\documentclass[twoside,twocolumn,english]{revtex4}
\usepackage[T1]{fontenc}
\usepackage[latin2]{inputenc}
\usepackage{float}
\usepackage{graphicx}

\makeatletter
\@ifundefined{textcolor}{}
{%
 \definecolor{BLACK}{gray}{0}
 \definecolor{WHITE}{gray}{1}
 \definecolor{RED}{rgb}{1,0,0}
 \definecolor{GREEN}{rgb}{0,1,0}
 \definecolor{BLUE}{rgb}{0,0,1}
 \definecolor{CYAN}{cmyk}{1,0,0,0}
 \definecolor{MAGENTA}{cmyk}{0,1,0,0}
 \definecolor{YELLOW}{cmyk}{0,0,1,0}
 }

\makeatother

\usepackage{babel}

\begin{document}

\title{Unsuitability of the moving light clock system for the Lorentz factor
derivation}

\author{Tomasz T. Wa\l{}ek}

\email{tomasz.walek@polsl.pl}

\affiliation{Department of Environmental and Safety Management, Silesian University
of Technology, 66 De Gaulle St., 41-800 Zabrze, Poland}
\begin{abstract}
The moving light clock system was analyzed with respect to the orientation
of the wavefront of the light pulse observed in the moving and stationary
frames of reference. The plane wavefront of the light pulse was oriented
horizontally in both the frames. The wavefront observed in the stationary
frame was not perpendicular to the direction of the light pulse propagation.
This showed different characteristics of the light pulse than that
assumed in the Lorentz factor derivation. According to the horizontal
orientation of the wavefront, velocity $c$ was determined as the
vertical component of the light pulse motion observed in the stationary
frame. Application of this velocity distribution in the Lorentz factor
derivation showed the same travel time for the light pulse observed
in the moving and stationary frames of reference. The moving light
clock system was therefore found to be unsuitable for the Lorentz
factor derivation and illustration of time dilation, and shown to
illustrate the relativity of the observation of light rather than
the relativity of time.
\end{abstract}



\maketitle

\section{Introduction}

The Lorentz factor plays a fundamental role in relativistic calculations.
It can be derived in several ways \cite{key-01,key-02,key-03,key-04,key-05,key-06,key-07,key-08,key-09,key-10,key-11,key-12,key-13,key-14},
nevertheless, the method of the moving light clock \cite{key-15}
is used in many textbooks and relativity courses. Therefore, it is
important to analyze all aspects of the applicability of this method.

\begin{figure}
\begin{centering}
\includegraphics[width=0.98\columnwidth]{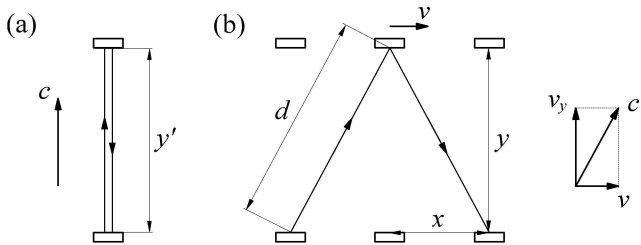} 
\par\end{centering}

\caption{\label{fig:Light_clock}The moving light clock observed in the moving
(a) and stationary (b) frame of reference.}

\end{figure}

Two frames of reference are considered in this derivation. Frame $S$
is stationary and frame $S'$ moves along the $x$-axis of frame $S$
with uniform velocity $v$. The light clock is arranged vertically
in frame $S'$ (Fig. \ref{fig:Light_clock}a).%
 It consists of two mirrors between which the light pulse emitted
in frame $S'$ travels in vacuum back and forth. An observer in frame
$S'$ observes the distance traversed by the light pulse between the
mirrors as $y'$ and the travel time as $t'$. An observer in frame
$S$ (Fig. \ref{fig:Light_clock}b) observes the distance as $d$
and the time as $t$. According to the postulate of invariance of
the speed of light, the light pulse travels in vacuum with velocity
$c$ regardless of the frame of reference. Thus, distance $y'$ can
be expressed as $ct'$ and distance $d$ as $ct$. Since distance
$d$ is greater than $y'$ then $ct>ct'$ and hence $t>t'$. This
means that the time in which the light pulse travels the distance
between the mirrors is longer in frame $S$ than in frame $S'$. By
equating distances $y'$ and $y$ we have $ct'=\sqrt{(ct)^{2}-(vt)^{2}}$
and we can calculate the Lorentz factor $\gamma=1/\sqrt{1-v^{2}/c^{2}}$.

This paper is intended to analyze the moving light clock system in
relation to the orientation of the wavefront of the light pulse observed
in the moving and stationary frames of reference.

\section{Orientation of the wavefront of the light pulse in the moving light
clock system}

The light pulse applied in the moving light clock system travels the
distance between the mirrors without dispersion and remains unchanged
in form. It means that the light of the light pulse is collimated.
Therefore, the light pulse traveling between the mirrors in vacuum
can be characterized by a plane wavefront which is perpendicular to
the wave normal and to the direction of the light pulse propagation.
This characteristic of the light corresponds to a laser beam and hence
the light pulse applied in the light clock is usually described as
a laser pulse. According to the principle of relativity, this characteristics
of the light pulse must be valid in any inertial reference frame.

\begin{figure}
\begin{centering}
\includegraphics[width=0.64\columnwidth]{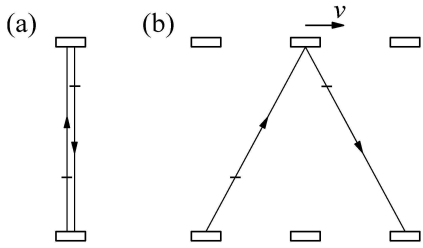} 
\par\end{centering}

\caption{\label{fig:Light_clock-wavefronts}The orientation of the wavefront
of the light pulse in the moving light clock system observed in the
moving (a) and stationary (b) frame of reference.}

\end{figure}

The light pulse observed in frame $S'$ travels vertically between
the mirrors and its wavefront is parallel to the surface of the mirrors
(Fig. \ref{fig:Light_clock-wavefronts}a).%
 The horizontally oriented wavefront will also be observed horizontally
in frame $S$ (Fig. \ref{fig:Light_clock-wavefronts}b), because the
speed of relative motion of frame $S'$ is zero in the direction $y$.
Hence, coordinates $y'$ and $y$ of any point of the wavefront must
have the same values as observed in both the reference frames. From
the same reason, the horizontal orientation of the mirrors is not
influenced by the motion of frame $S'$. (See Appendix for additional
discussion of the orientation of the wavefront.)

\begin{figure}
\begin{centering}
\includegraphics[width=0.65\columnwidth]{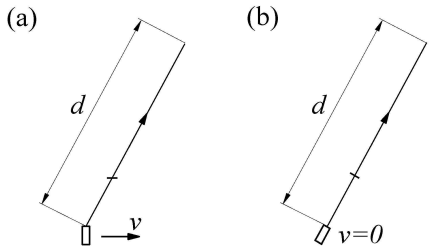} 
\par\end{centering}

\caption{\label{fig:Wavefronts-difference}The orientation of the wavefront
of the light pulse emitted in the moving light clock system (a) and
in the stationary frame of reference (b), as observed in the stationary
frame.}

\end{figure}

It can be seen that the orientation of the wavefront of the light
pulse observed in frame $S$ is not perpendicular to the direction
of the light pulse propagation (Fig. \ref{fig:Wavefronts-difference}a).
The light pulse therefore does not correspond to the characteristics
of laser light traveling in vacuum. Such a situation, where the wavefront
of the light pulse is perpendicular to the direction of propagation,
occurs for the light pulse emitted in the stationary frame of reference
along path $d$, independently of the moving light clock system (Fig.
\ref{fig:Wavefronts-difference}b).%
 In both the cases the orientation of the wavefront is determined
by the orientation of the emitter during emission. The distances traveled
in both cases are the same, but the light pulses cannot be considered
identical because of the different wavefront orientations.

This shows the complex character of the light pulse motion observed
in the moving light clock system in frame $S$. In this case the motion
of the light pulse cannot be treated as an independent motion. It
is a visual summation of two independent motions -- the light pulse
moving in relation to frame $S'$ and frame $S'$ moving in relation
to frame $S$. Therefore, it cannot be said that the light pulse travels
distance $d$ in frame $S$ with velocity $c$, as it is assumed in
the Lorentz factor derivation, because the light pulse observed in
frame $S$ does not correspond to the characteristics of the light
pulse traveling in vacuum. We can only state that the motion of the
light pulse and the motion of frame $S'$ form in frame $S$ an image
of the light pulse characterized by the wavefront oriented not perpendicularly
to the direction of propagation, and moving along path $d$.

\section{Velocity distribution in the moving light clock system}

Since the plane wavefront of the light pulse traveling in vacuum is
perpendicular to the direction of propagation, and the light pulse
propagates in vacuum with velocity $c$, this means that velocity
$c$ is perpendicular to the wavefront of the light pulse. It shows
that the orientation of velocity $c$ is closely related to the orientation
of the wavefront. Because of the principle of relativity, this relation
must be valid in any inertial reference frame.

\begin{figure}
\begin{centering}
\includegraphics[width=0.77\columnwidth]{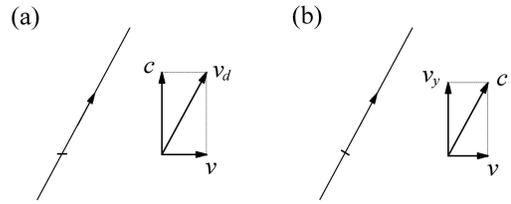} 
\par\end{centering}

\caption{\label{fig:Velocity_distributions}The velocity distribution of the
light pulse emitted in the moving light clock system (a) and in the
stationary frame of reference (b), as observed in the stationary frame.}

\end{figure}

For the moving light clock system observed in frame $S$, this means
that velocity $c$ is the vertical component of the motion of the
light pulse (Fig. \ref{fig:Velocity_distributions}a).%
 The component velocity $v$ is in this case not related to the motion
of the light pulse. It is a result of the movement of frame $S'$
in relation to frame $S$ only, and any change in velocity of frame
$S'$ will affect it.

This velocity distribution differs from the assumptions of the Lorentz
factor derivation, where velocity $c$ is supposed to occur along
path $d$ (Fig. \ref{fig:Light_clock}b). Such a situation, where
velocity $c$ is oriented along path $d$ and is perpendicular to
the wavefront of the light pulse, occurs for the light pulse emitted
in the stationary frame of reference, independently of the moving
light clock system (Fig. \ref{fig:Velocity_distributions}b). In this
case the component velocity $v$ is related to the motion of the light
pulse only -- no movement of any reference frame in relation to frame
$S$ can influence it.

\section{Discussion}

The analysis of the orientation of the wavefront of the light pulse
in the moving light clock system shows that the horizontally oriented
wavefront observed in frame $S'$ will also be observed horizontally
in frame $S$ (Fig. \ref{fig:Light_clock-wavefronts}). The light
pulse observed in the stationary frame is characterized by the wavefront
oriented not perpendicular to the direction of propagation. This shows
the complex character of the light pulse motion observed in frame
$S$. The motion observed in frame $S$ along path $d$ is a visual
summation of the motions of the light pulse in relation to frame $S'$
and frame $S'$ in relation to frame $S$.

According to the horizontal orientation of the wavefront in frame
$S$ it can be seen, that the velocity distribution with vector $c$
oriented vertically (Fig. \ref{fig:Velocity_distributions}a) should
be applied in the Lorentz factor derivation. In this case we have
$ct'=\sqrt{(v_{d}t)^{2}-(vt)^{2}}$, which leads to $t'=t$. It means
that the travel time of the light pulse is the same in the moving
and stationary frames of reference. The occurrence of velocity $v_{d}$
greater than $c$ does not mean exceeding the speed of light by the
light pulse in this case, because the motion observed in frame $S$
along path $d$ is not an independent motion of the light pulse.

Both the complex character of the observation of the light pulse in
frame $S$ and the velocity distribution resulting from the horizontal
orientation of the wavefront of the light pulse show that the moving
light clock system is unsuitable for the Lorentz factor derivation
and illustration of time dilation. The analysis presented in this
paper suggests that the moving light clock system presents an argument
for the simple emission theory rather than the special relativity
theory, and illustrates the relativity of the observation of light
rather than the relativity of time.

\section{Conclusions}

The plane wavefront of the light pulse in the moving light clock system
is oriented horizontally both in the moving and stationary frames
of reference. The light pulse observed in the stationary frame is
characterized by the wavefront oriented not perpendicular to the direction
of propagation. It shows different characteristics of the light pulse
than that assumed in the Lorentz factor derivation.

The horizontal orientation of the wavefront observed in the stationary
frame of reference allows to determine the velocity distribution of
the light pulse with vertically oriented velocity $c$. Application
of this velocity distribution in the Lorentz factor derivation shows
that the travel time of the light pulse is the same in the moving
and stationary frames of reference.

The analysis of the moving light clock system in relation to the orientation
of the wavefront of the light pulse shows that the moving light clock
system is unsuitable for the Lorentz factor derivation and illustration
of time dilation. The moving light clock system illustrates the relativity
of the observation of light rather than the relativity of time.\appendix*

\section{Horizontal orientation of the wavefront}

The horizontal orientation of the wavefront observed both in the moving
and stationary frames of reference can also be verified as follows.

\begin{figure}[H]
\begin{centering}
\includegraphics[width=1\columnwidth]{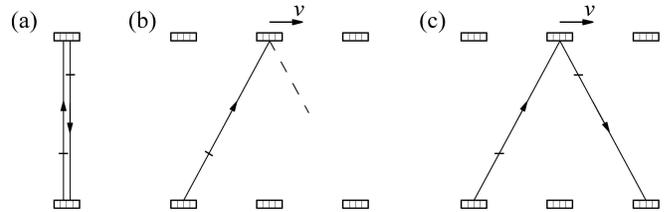} 
\par\end{centering}

\caption{\label{fig:Selective_reflection}The moving light clock with selective
reflection mirrors, as observed in frame $S$: (a) the light clock
remaining stationary in relation to frame $S$; (b) the light clock
moving in relation to frame $S$, with the wavefront oriented not
horizontally; (c) the light clock moving in relation to frame $S$,
with the wavefront oriented horizontally.}

\end{figure}

The moving light clock system can be modified so that selective reflection
mirrors are applied, which reflect only the light with a wavefront
parallel to the mirror surface. When the modified light clock remains
stationary in relation to frame $S$, the observers in both the reference
frames agree that the plane wave of the light pulse reaches the mirror
parallel to its surface (Fig. \ref{fig:Selective_reflection}a).%
 In this case the reflection will occur and the observers will observe
the light pulse traveling back and forth between the mirrors. However,
when the light clock starts moving along the $x$-axis of frame $S$,
we can consider two possibilities with regard to the observation of
the wavefront in frame $S$ -- the first one with the slanted wavefront
(Fig. \ref{fig:Selective_reflection}b), and the second one with the
wavefront oriented horizontally (Fig. \ref{fig:Selective_reflection}c).

In the first case, the wavefront of the light pulse reaches the mirror
in frame $S$ not parallel to its surface (Fig. \ref{fig:Selective_reflection}b),
and consequently the reflection cannot occur. This means that any
motion of the light clock along the $x$-axis, even with only minimal
velocity, should exclude the reflection of the light pulse in frame
$S$. At the same time, the observer in frame $S'$ will observe the
light pulse as reflecting between the mirrors, because the angle of
incidence of the light pulse observed in frame $S'$ cannot be influenced
by any motion of the light clock in relation to frame $S$. Thus,
the assumption of the non-horizontal orientation of the wavefront
leads to a paradoxical situation where the reflection of the light
pulse can be observed in one frame of reference but not in the other.

The exclusion of reflection observed in frame $S$ does not occur
in the second case, in which the horizontally oriented wavefront reaches
the mirror parallel to the surface (Fig. \ref{fig:Selective_reflection}c).
In this case the motion of the light pulse and its reflection can
be observed in both the reference frames. In fact, the reflection
takes place in frame $S'$, and in frame $S$ it is only an observation.

Therefore, the plane wavefront observed in the moving light clock
system is shown to be oriented horizontally in both the moving and
stationary frames of reference.


\begin{thebibliography}{15}
\bibitem{key-01}Brehme, R.W.: A Geometric Representation of Galilean
and Lorentz Transformations. Am. J. Phys. 30, 489--496 (1962)

\bibitem{key-02}Einstein, A.: Relativity: The Special and General
Theory, pp. 131--138. Three Rivers Press, New York (1995)

\bibitem{key-03}de Felice, F., Sigalotti, L.D.G., Mejias, A.: Lorentz
transformations and complex space-time functions. Chaos, Solitons
\& Fractals 21, 573--578 (2004)

\bibitem{key-04}Fowles, G.R.: Self-inverse form of the Lorentz transformation.
Am. J. Phys. 45, 675--676 (1977)

\bibitem{key-05}Harvey, A.L.: Lorentz Transformation. Am. J. Phys.
36, 901--904 (1968)

\bibitem{key-06}Lee, A.R., Kalotas, T.M.: Lorentz transformations
from the first postulate. Am. J. Phys. 43, 434--437 (1975)

\bibitem{key-07}L\'{e}vy-Leblond, J-M.: One more derivation of the
Lorentz transformation. Am. J. Phys. 44, 271--277 (1976)

\bibitem{key-08}Macdonald, A.: Derivation of the Lorentz transformation.
Am. J. Phys. 49, 493 (1981)

\bibitem{key-09}Nadeau, G.: The Lorentz-Einstein Transformation Obtained
by a Vector Method. Am. J. Phys. 30, 602--603 (1962)

\bibitem{key-10}Pfleiderer, J.: Lorentz Transformation Derived from
First-Order Experiments. Am. J. Phys. 37, 1131--1134 (1969)

\bibitem{key-11}Schwartz, H.M.: Deduction of the general Lorentz
transformations from a set of necessary assumptions. Am. J. Phys.
52, 346--350 (1984)

\bibitem{key-12}Sen, A.: How Galileo could have derived the special
theory of relativity. Am. J. Phys. 62, 157--162 (1994)

\bibitem{key-13}Sigalotti, L.D.G, Mejias, A.: The golden ratio in
special relativity. Chaos, Solitons \& Fractals 30, 521--524 (2006)

\bibitem{key-14}Srivastava, A.M.: Invariant speed in special relativity.
Am. J. Phys. 49, 504--505 (1981)

\bibitem{key-15}Feynman, R.P., Leighton, R.B., Sands, M.: The Feynman
Lectures on Physics, pp. 15-5--15-6. Addison-Wesley, Reading, Massachusetts
(1963)
\end{thebibliography}
\end{document}